\documentclass[fleqn,10pt]{SelfArx}
\usepackage{lipsum}

\definecolor{color1}{RGB}{0,0,90} % Color of the article title and sections
\definecolor{color2}{RGB}{0,20,20} % Color of the boxes behind the abstract an

\usepackage{hyperref} % Required for hyperlinks
\hypersetup{hidelinks,colorlinks,breaklinks=true,urlcolor=color2,citecolor=color1,linkcolor=color1,bookmarksopen=false,pdftitle={Title},pdfauthor={Author}}

 \renewcommand{\vec}[1]{\mbox{\boldmath $#1$}}

 \def\gsim{\lower.4ex\hbox{$\;\buildrel >\over{\scriptstyle\sim}\;$}}

 \def\aap{A\&A}
 \def\apj{ApJ}

 \def\mnras{MNRAS}

\JournalInfo{Astronomy Letters, 2017, Vol. 43, No. 9} % Journal information
\Archive{}
\PaperTitle{
Double Hall instability: A catalyzer of magnetic energy release
}
\Authors{L.~L.~Kitchatinov\textsuperscript{1,2}*}
\affiliation{\textsuperscript{1}\textit{Institute for Solar-Terrestrial Physics, Lermontov Str. 126A, Irkutsk, 664033, Russia}}
\affiliation{\textsuperscript{2}\textit{
Pulkovo Astronomical Observatory, Pulkovskoe Sh. 65, St. Petersburg, 196140, Russia
}}
\affiliation{*\textbf{E-mail}: kit@iszf.irk.ru}

\Keywords{instabilities - magnetic fields - stars: neutron}
\newcommand{\keywordname}{Keywords}

\Abstract{A pictorial explanation for shear-Hall instability is suggested and shows that the shear flow is not necessary for the instability because its role can be played by the Hall effect of an inhomogeneous background magnetic field. Linear stability analysis for a simple model of magnetic field varying periodically in space confirms such a \lq double Hall' instability. Numerical computations show a considerable increase in Ohmic dissipation rate at the nonlinear stage of instability development. Field dissipation has a spiky character associated with magnetic reconnection in current sheets and X-points. Double Hall instability can be significant for magnetic field dissipation in neutron star crusts and, possibly, in the solar corona.}
\begin{document}

\flushbottom % Makes all text pages the same height

\maketitle % Print the title and abstract box

%\tableofcontents % Print the contents section

\thispagestyle{empty} % Removes page numbering from the first page

%----------------------------------------------------------------------------------------
%	ARTICLE CONTENTS
%----------------------------------------------------------------------------------------
%%%%%%%%%%%%%%%%%%%%%%%%%%%%%%%%%%%%%%%%%%%%%%%%%%%%%%%%%%%%%%%%%%%%%%%%%%%%%
\section{Introduction} % The \section*{} command stops section numbering
%%%%%%%%%%%%%%%%%%%%%%%%%%%%%%%%%%%%%%%%%%%%%%%%%%%%%%%%%%%%%%%%%%%%%%%%%%%%%
Magnetic field dynamics with the Hall effect, also known as the Electron-MHD (Gour\-go\-u\-li\-a\-tos \& Hollerbach 2016), is studied mainly in relation with neutron star magnetism. Jones (1988) was probably the first to note that the characteristic time $\sim 10^7$ yr of pulsars' magnetic field decay, which is short compared to Ohmic decay time, can be related to the Hall drift of the fields. Later on, Goldreich \& Reisenegger (1992) analysed various effects in pulsars' magnetic field dynamics and confirmed the primary importance of the Hall effect for neutron star crusts. They supposed that the Hall drift initiates fragmentation of magnetic field scales to drive the magnetic energy cascade towards small scales of Ohmic dissipation. The scenario was required because the Hall effect does not change the magnetic energy by itself and can influence field dissipation as a \lq catalyzer' of the Ohmic decay only. An instability initiating the energy cascade by Hall turbulence has not been specified however.

Other possibilities for decreasing the magnetic fields of pulsars include magnetic field screening by accreting matter (Bisnovatyi-Kogan \& Komberg 1974; Bisnovatyi-Kogan 2016) or enhancement of Ohmic decay by accretion (Urpin \& Geppert 1995). The possibilities do not apply, however, to a small group of isolated neutron stars. The radiation power of this type of pulsar is in all probability maintained by the release of magnetic energy (Heyl \& Kulkarni 1998; Harding 2013). Also, the importance of accretion does not abate the significance of the Hall effect.

A Hall-induced decrease in the field scale can occur for two reasons: lack of equilibrium and/or instability. The Hall equilibrium of the magnetic filed $\vec B$ is defined as a state with a potential vector ${\vec B}\times({\vec\nabla}\times{\vec B})$ for which the contribution of the Hall effect in the induction equation vanishes.
Non-equilibrium dynamics of magnetic fields has been studied in a series of -- mainly numerical -- works (Shalybkov \& Urpin 1997; Urpin \& Shalybkov 1999; Hollerbach \& R\"udiger 2002; Cumming et al. 2004; Kojima \& Kisaka 2012; Marchant et al. 2014). Global oscillations with the participation of relatively small-scale modes of magnetic field were found. A considerable amplification of Ohmic decay, however, did not occur. Even a regime with an increase in dipolar magnetic momentum with time was found (Gourgouliatos \& Kumming 2015). In some cases, \lq numerical' instabilities were met which did not allow computations for large Hall parameters.

Hall equilibria can, however, be unstable. In these cases, small deviations from an equilibrium grow exponentially with time. This paper concerns such an instability. The consideration starts with a discussion of the shear-Hall instability (R\"udiger \& Hollerbach 2004), which shows that the effect of the shear flow in the instability can be replaced by the Hall effect of an inhomogeneous background field. The instability is then developing in the vicinity of the local maxima of the field strength. This is probably the same instability as discovered by Rheinhardt \& Geppert (2002). This paper formulates the necessary condition for its development. A pictorial explanation for the instability is suggested which allows an estimation of its expected parameters. Linear stability analysis for a particular model of an inhomogeneous background field confirms these estimations. The instability is fast: growth rates of unstable disturbances exceed the rate of Ohmic dissipation of the background field by far. The unstable disturbances are large-scaled along the background field but vary on small spatial scales across  this field.

Nonlinear computations of the instability show formation of the current sheets accom\-pa\-ni\-ed by an increase in the rate of Ohmic dissipation. The length of the current sheets decreases with time but the dissipation rate increases simultaneously. The rate attains its maximum value at the instant of the current sheet transformation into an X-point. Reconnection in the X-point changes the field structure totally. The new field distribution can, however, be unstable as well and the events of the current sheet formation with spikes of energy release repeat. The number of repetitions depends on the Hall number and increases with this number. After a sufficiently long time, the field distribution approaches a stable Hall equilibrium that terminates the instability-induced release of magnetic energy.

The paper concludes with a discussion of possible applications of its results.
%%%%%%%%%%%%%%%%%%%%%%%%%%%%%%%%%%%%%%%%%%%%%%%%%%%%%%%%%%%%%%%%%%%%%%%%%%%%%
\section{Preliminary estimations}
%%%%%%%%%%%%%%%%%%%%%%%%%%%%%%%%%%%%%%%%%%%%%%%%%%%%%%%%%%%%%%%%%%%%%%%%%%%%%
The Hall drift can be significant for magnetic fields of sufficient strength when the cyclotron frequency of the main carriers of the electric current -- electrons -- is not small compared to the frequency of their collisions with particles of other species. The dimensionless Hall parameter used below is in fact the ratio of these two frequencies.

The induction equation for the neutron star crusts with allowance for the Hall effect reads (Goldreich \& Reisenegger 1992)
\begin{eqnarray}
    &&\frac{\partial{\vec B}}{\partial t}\ = - {\vec\nabla}\times\left(
    \beta\left({\vec\nabla}\times{\vec B}\right)\times{\vec B}
    + \eta{\vec\nabla}\times{\vec B}\right)\, ,
    \nonumber \\
    && \beta = \frac{c}{4\pi n e},\ \ \  \eta = \frac{c^2}{4\pi\sigma} .
    \label{1}
\end{eqnarray}
In this equation, $n$ is the electron density, $e$ is the elementary electric charge, $\sigma$ is the conductivity, and other standard notations are used. This paper concerns the stability of Hall equilibria, i.e., of such distributions of the field for which the contribution of the Hall effect in Eq.\,(\ref{1}) is zero: ${\vec\nabla}\times\left(
\beta\left({\vec\nabla}\times{\vec B}\right)\times{\vec B}\right) = 0$. The consideration restricts itself by a simple case of spatially uniform coefficients $\eta$ and $\beta$, and the background field only one $z$-component of which in a Cartesian coordinate system $(x,y,z)$ differs from zero.

For the simplest case of a uniform field ${\vec B} = {\vec e}_z B_0$ (${\vec e}_z$ is the unit vector along the $z$-axis), the magnetic disturbances ${\vec b} \sim \mathrm{exp}\left(-\mathrm{i}\omega t + \mathrm{i}kz\right)$ represent the well-known helicoidal oscillations:
\begin{equation}
    \omega = \pm\beta k^2 B_0 - \mathrm{i}k^2\eta .
    \label{2}
\end{equation}
It is important for what follows that the magnetic disturbances in these oscillations rotate around the background magnetic field:
\begin{eqnarray}
    b_x(z,t) &=& b_0(z)\mathrm{e}^{-\eta k^2 t}\cos(\beta k^2 B_0 t + \phi_0)\, ,
    \nonumber \\
    b_y(z,t) &=& b_0(z)\mathrm{e}^{-\eta k^2 t}\sin(\beta k^2 B_0 t + \phi_0),
    \label{3}
\end{eqnarray}
where $b_0$ is the initial amplitude and $\phi_0$ is the initial phase of the disturbances. The rotation is clockwise if seen in the background field direction\footnote{For the disturbances with the wave vector not aligned with the background field, the rotation proceeds around the axis defined by the wave vector and its sense depends on the sign of the background field projection on this axis.}.

The uniform field whose disturbances obey equation (\ref{1}) is obviously stable.
How\-ev\-er, the Hall effect leads to an instability even for a uniform field if the matter in which the field is frozen is a fluid undergoing a shear flow (R\"udiger \& Hollerbach 2004). In this case, the term ${\vec\nabla}\times\left({\vec V}\times{\vec B}\right)$ should de added to the right-hand side of the equation (\ref{1}). For the shear flow
\begin{equation}
    {\vec V} = -{\vec e}_y S x
    \label{4}
\end{equation}
(${\vec e}_y$ is the unit vector along the $y$-axis) with given and spatially uniform vorticity $S$, instead of Eq.\,(\ref{2}) one finds
\begin{equation}
    \omega = \pm\mathrm{i}\sqrt{\beta k^2B_0( S - \beta k^2 B_0)} -
    \mathrm{i}\eta k^2 .
    \label{5}
\end{equation}
The instability is possible for a positive value of the expression under the square root in equation (\ref{5}). It takes place if the inequality
\begin{equation}
    \frac{S}{\beta B_0 k^2} > 1 + \frac{\eta^2}{\beta^2 B_0^2}
    \label{6}
\end{equation}
is satisfied. The instability is present for sufficiently long waves (sufficiently small $k$) but only if the vorticity $S$ and the magnetic field $B_0$ are simultaneously either positive or negative.

\begin{figure}\centering
\includegraphics[width=\linewidth]{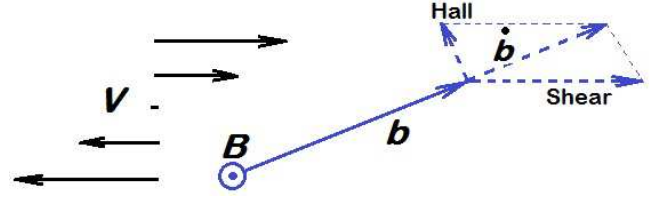}
\caption{An illustration for the origin of shear-Hall in\-sta\-bi\-li\-ty.
    The shear-flow profile is shown on the left. The background magnetic field $\vec B$ is normal to the figure plane. An exponential growth of magnetic disturbances $\vec b$ occurs when the helicoidal rotation due to the Hall effect and the deformation of the disturbances by the shear flow act in opposite directions, i.e., when ${\vec B}\cdot({\vec\nabla}\times{\vec V}) < 0$.}
    \label{f1}
\end{figure}

This finding can be given a clear pictorial explanation (Fig.\,\ref{f1}). The Hall effect gives the helicoidal rotation of the magnetic disturbances $\vec b$ and the shear flow deforms them. The rate of change $\dot{\vec b}$ of the disturbances due to either of the effects is proportional to the disturbance amplitude. The disturbances grow exponentially when the background field $\vec B$ and the vorticity ${\vec\nabla}\times{\vec V}$ are counter-aligned.

Equation (\ref{1}) governs the magnetic field frozen-in to the \lq electron fluid'  carrying the electric current in the solid crust. Apart from the Ohmic diffusion, the equation describes the filed transport with the effective velocity
\begin{equation}
    \tilde{\vec V} = -{\vec j}/(ne),
    \label{7}
\end{equation}
where ${\vec j} = c({\vec\nabla}\times{\vec B})/(4\pi)$ is the current density (the minus sign in the equation for the effective velocity corresponds to the negative electron charge).

The effective velocity (\ref{7}) can replace the shear flow in the shear-Hall instability and lead to an instability driven exclusively by the Hall effect.
It is in this sense that the title of this paper speaks of {\sl double} Hall instability.

According to Fig.\,\ref{f1}, double Hall instability can be expected only if the curl of the effective velocity is counter-aligned with the background magnetic field:
\begin{equation}
    {\vec B}\cdot\left({\vec\nabla}\times({\vec\nabla}\times{\vec B})\right) > 0 .
    \label{8}
\end{equation}
It can be seen easily that this condition is satisfied in the vicinity of the local maxima of field strength where ${\vec B} \simeq {\vec e}_z B_0 (1 - x^2/L_x^2)(1 - y^2/L_y^2)$ independently of the sign of $B_0$. The condition is satisfied everywhere for the field distribution
\begin{equation}
    {\vec B} = {\vec e}_z B_0 \cos(\kappa x) ,
    \label{9}
\end{equation}
by which all the following computations are confined. Apart from condition (\ref{8}), such a choice is motivated by the assumption repeatedly met in the literature that the Hall equilibria which are not changed by Ohmic diffusion are stable. The distribution of Eq.\,(\ref{9}) is a particular case of such an equilibrium but it is not stable.

Let us estimate the spatial scale of the supposed instability. At the local maxima of field (\ref{9}) at the points $x = 2\pi m/\kappa$ ($m$ is an integer), the vorticity of the effective velocity is $\tilde{S} = \beta{\vec e}_z\cdot \left({\vec\nabla}\times({\vec\nabla}\times{\vec B})\right) = \beta B_0\kappa^2$. A substitution of this expression in Eq.\,(\ref{6}) shows that the instability can be expected for $k < \kappa$. At first glance, this excludes an enhancement of the Ohmic decay: the instability is large-scaled along the background field lines. Inhomogeneity of the conditions for development of the instability can, however, result in small scales of the unstable disturbances across the background field.  The computations to follow confirm this expectation.
%%%%%%%%%%%%%%%%%%%%%%%%%%%%%%%%%%%%%%%%%%%%%%%%%%%%%%%%%%%%%%%%%%%%%%%%%%%%%
\section{Linear stability analysis}
%%%%%%%%%%%%%%%%%%%%%%%%%%%%%%%%%%%%%%%%%%%%%%%%%%%%%%%%%%%%%%%%%%%%%%%%%%%%%
\subsection{Equations}
%%%%%%%%%%%%%%%%%%%%%%%%%%%%%%%%%%%%%%%%%%%%%%%%%%%%%%%%%%%%%%%%%%%%%%%%%%%%%
We proceed by considering the stability of magnetic field (\ref{9}) to small disturbances. A decrease in the background field due to finite conductivity is neglected. Only rapidly growing disturbances with the growth rates $\gamma \gg \kappa^2\eta$ can, therefore, be significant. The infinitesimal disturbances $\vec b$ are assumed to not vary along the $y$-axis. It is convenient to use the following representation for such 2D disturbances
\begin{equation}
    {\vec b}(x,z) = {\vec e}_y b(x,z)
    + {\vec\nabla}\times\left({\vec e}_y a(x,z)\right).
    \label{10}
\end{equation}
The first and the second terms on the right-hand side of Eq.\,(\ref{10}) will be named the toroidal and poloidal fields, respectively.

The variables are normalized to dimensionless units. Time is measured in the diffusive units of $(\kappa^2\eta)^{-1}$ and distance -- in units of $\kappa^{-1}$. The same notations as before are kept for such normalized time and spatial coordinates. Linearization of Eq.\,(\ref{1}) in small deviations from the background field (\ref{9}) leads to the following equation system
\begin{eqnarray}
     \frac{\partial b}{\partial t} &=& 2R_\mathrm{H}\cos(x)\left(
     \frac{\partial (\Delta a)}{\partial z} + \frac{\partial a}{\partial z}\right)
     + \Delta b\ ,
     \nonumber \\
     \frac{\partial a}{\partial t} &=&-2R_\mathrm{H} \cos(x)
     \frac{\partial b}{\partial z} + \Delta a\ ,
     \label{11}
\end{eqnarray}
where $\Delta = \partial^2/\partial x^2 + \partial^2/\partial z^2$ is the 2D Laplacian and
\begin{equation}
    R_\mathrm{H} = \frac{\beta B_0}{2\eta} = \frac{\sigma B_0}{2 c n e}
    \label{12}
\end{equation}
is the Hall parameter.

The coefficients in Eqs\,(\ref{11}) do not depend on $z$ and time and vary periodically with $x$. These allow a search for the solution in the form
\begin{eqnarray}
    a = \mathrm{e}^{\gamma t}\sin(\hat{k}z) \left(
    \sum\limits_{n=0}^{N}a^c_n\cos(nx) +
    \sum\limits_{n=1}^{N}a^s_n\sin(nx)\right),&&
    \nonumber \\
    b = \mathrm{e}^{\gamma t}\cos(\hat{k}z) \left(
    \sum\limits_{n=0}^{N}b^c_n\cos(nx) +
    \sum\limits_{n=1}^{N}b^s_n\sin(nx)\right).&&
    \label{13}
\end{eqnarray}
Substitution of Eqs\,(\ref{13}) into (\ref{11}) leads to the eigenvalue problem for a system of algebraic equations.

Complex arithmetics conventional for the eigenvalue problems is not used in Eqs\,(\ref{13}) yet. This allows one to see that the complete system of equations splits into four independent subsystems governing eigenmodes of different spatial structure.
There are symmetric in coordinate $x$ S-modes and antisymmetric A-modes represented on the right-hand side of Eqs\,(\ref{13}) by the first and the second sums, respectively. The S-modes decouple further in two independent groups of modes combining the coefficients $(a^c_{2n},b^c_{2n+1})$ or $(a^c_{2n+1},b^c_{2n})$ with even or odd values of their subscripts. These two groups of modes are notated as S1 and S2, respectively. The modes A1 and A2 can be defined similarly.

The eigenvalue problem was solved numerically. For \\ $R_\mathrm{H} \leq 1000$, the series expansions in Eqs (\ref{13}) converge rapidly and the results for $N = 200$ and $N = 300$ are practically identical. Only such resolution-independent results are discussed below.

To conclude the mathematical formulation, it may be no\-ted that the problem allows an analytical solution in the perfect conductivity limit. In this limit, the last terms on the right-hand sides of Eqs\,(\ref{11}) can be omitted and differentiation of the second one of these equations over time reduces the problem to a single equation for the potential $a$ of the poloidal field. A curious and not finally understood property of the perfect conductivity limit is the absence of the instability which is confidently detected for whatever large but finite conductivity. This property is, however, of pure academic interest because the considered instability can be significant only in relation with an expected amplification of the Ohmic dissipation.  The dissipation is certainly not possible in the perfect conductivity limit which is therefore not discussed any further.

%%%%%%%%%%%%%%%%%%%%%%%%%%%%%%%%%%%%%%%%%%%%%%%%%%%%%%%%%%%%%%%%%%%%%%%%%%%%%%
\subsection{Results}
%%%%%%%%%%%%%%%%%%%%%%%%%%%%%%%%%%%%%%%%%%%%%%%%%%%%%%%%%%%%%%%%%%%%%%%%%%%%%%
All eigenmodes except S1 are stable. They represent the decaying helicoidal oscillations with finite imaginary and negative real parts of their corresponding eigenvalues. Among the multiplicity of S1 modes, there is one unstable. The unstable eigenmode is not oscillatory but grows steadily with time (a change of stability). This is consistent with the pictorial explanation of Fig.\,\ref{f1}.

\begin{figure}\centering
\includegraphics[width=7truecm]{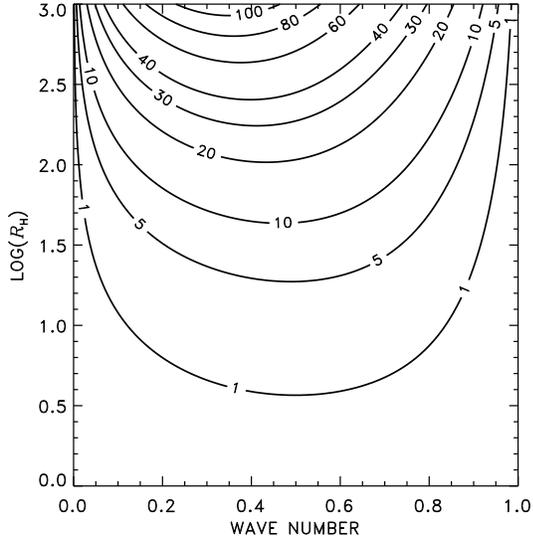}
\caption{Isolines of the growth rates of the double Hall instability on the
    plane of the Hall parameter (\ref{12}) and the (normalized) wave number $\hat{k} = k/\kappa$. The numbers in the isoline gaps show the growth rates in units of the rate $\eta\kappa^2$ \\ of the Ohmic decay of the background field.}
    \label{f2}
\end{figure}

Figure\,\ref{f2} shows the contour lines of the growth rates on the plane of the Hall parameter $R_\mathrm{H}$ and the wave number $\hat{k} = k/\kappa$. The instability is fast: the growth rate $\gamma$ increases with the Hall parameter and exceeds the rate of Ohmic dissipation by two orders of magnitude for $R_\mathrm{H} = 1000$.

\begin{figure}\centering
\includegraphics[width=\linewidth]{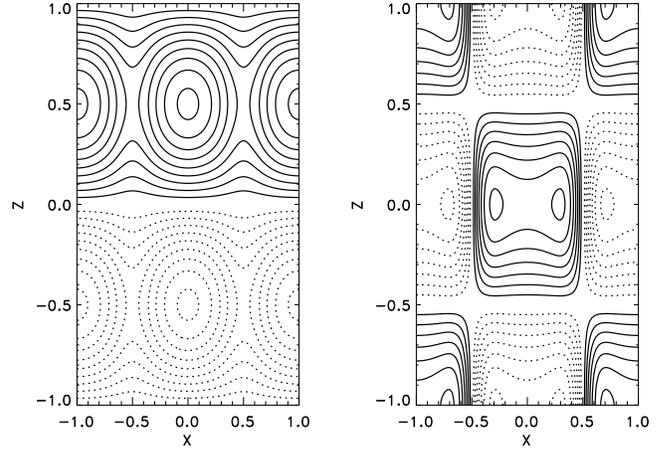}
 \caption{The pattern of unstable magnetic disturbance for the Hall
    parameter $R_\mathrm{H} = 10$ and the wave number $\hat{k} = 0.5$, for which the maximum growth rate $\gamma = 2.86$ is attained. {\sl Left panel:} the poloidal field lines in coordinates of $X = \pi x$ and $Z = \pi\hat{k}z$. {\sl Right panel:} isolines of the toroidal field $b$. Full (dotted) lines show the lines of clockwise (anti-clock- wise) circulation of the poloidal field vector and positive (negative) levels of the toroidal field.
    }
 \label{f3}
\end{figure}

\begin{figure}\centering
\includegraphics[width=\linewidth]{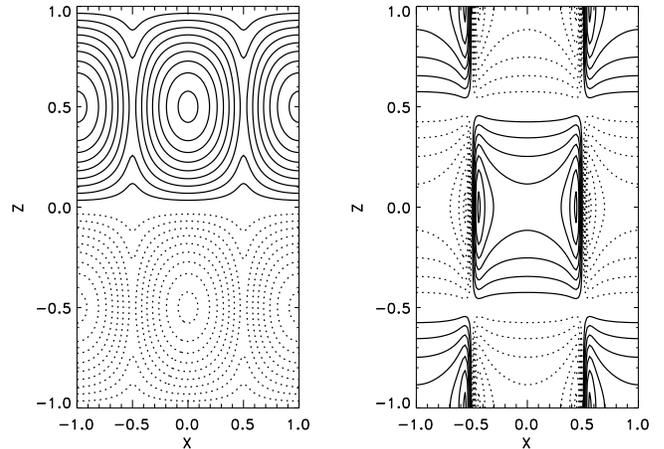}
 \caption{\small The same as in Fig.\,\ref{f3} but for the
    Hall parameter  $R_\mathrm{H} = 100$ ($\hat{k} = 0.434,\ \gamma = 19.5$).
    }
 \label{f4}
\end{figure}

As expected, the wave-lengths of unstable disturbances are long $\hat{k} < 1$. A fine structure is however present in the dimension across the background field. Figures \ref{f3} and \ref{f4} show the patterns of unstable disturbances on the plane of the coordinates $X = \pi x$ and $Z = \pi\hat{k}z$. The figures cover one period of the spatially periodic disturbances in either coordinate. The disturbances have the largest magnitude near the maximum of the background field at $X = 0$. A comparison of Figs \ref{f3} and \ref{f4} shows that the disturbance inhomogeneity across the background field steepens with increasing Hall parameter (with increasing strength of the background field). This indicates the possible enhancement of the Ohmic dissipation by the instability.

Note that the effective velocity (\ref{7}) for the toroidal field patterns of Figs \ref{f3} and \ref{f4} corresponds to a flow converging along the $X$-axis and diverging along the axis $Z$ near the points $(X,Z)=(0.5,-0.5)$ and $(-0.5,0.5)$. Such a flow pattern is typical of the current sheets. The possibility of the current sheets formation due to the Hall effect has been noted in several publications (Vainshtein et al. 2000; Reisenegger et al. 2007; Pons \& Geppert 2007, 2010; Marchant et al. 2014). A realisation of this possibility by the double Hall instability can only be shown by nonlinear computations to which we proceed.
%%%%%%%%%%%%%%%%%%%%%%%%%%%%%%%%%%%%%%%%%%%%%%%%%%%%%%%%%%%%%%%%%%%%%%%%%%%%%
\section{Nonlinear dynamics of the instability}
%%%%%%%%%%%%%%%%%%%%%%%%%%%%%%%%%%%%%%%%%%%%%%%%%%%%%%%%%%%%%%%%%%%%%%%%%%%%%
\subsection{Equations}
%%%%%%%%%%%%%%%%%%%%%%%%%%%%%%%%%%%%%%%%%%%%%%%%%%%%%%%%%%%%%%%%%%%%%%%%%%%%%
The magnetic field can be written as a superposition of its toroidal and poloidal parts,
\begin{equation}
    {\vec B}(x,z) = {\vec e}_y B(x,z) + {\vec\nabla}\times\left({\vec e}_y A(x,z)\right) ,
    \label{14}
\end{equation}
similar to Eq.\,(\ref{10}) for small disturbances. The same normalized variables as in the linear problem are used and the magnetic field is normalized to its initial amplitude $B_0$. A substitution of Eq.\,(\ref{14}) in (\ref{1}) leads to the system of two equations:
\begin{eqnarray}
    \frac{\partial B}{\partial t} &=& 2R_\mathrm{H} \left(
    \frac{\partial A}{\partial x}\frac{\partial(\Delta A)}{\partial z} -
    \frac{\partial A}{\partial z}\frac{\partial(\Delta A)}{\partial x}
    \right)
    + \Delta B\ ,
    \nonumber \\
    \frac{\partial A}{\partial t} &=& 2R_\mathrm{H} \left(
    \frac{\partial B}{\partial x}\frac{\partial A}{\partial z} -
    \frac{\partial B}{\partial z}\frac{\partial A}{\partial x}\right) +
    \Delta A\ ,
    \label{15}
\end{eqnarray}
where $R_\mathrm{H}$ is the Hall parameter (\ref{12}). The initial value problem for  Eqs\,(\ref{15}) was considered.

The initial field was prescribed as a superposition of the background field (\ref{9}) with a small addition of the unstable mode of the linear problem:
\begin{eqnarray}
    A &=& \sin(x) + \varepsilon \sin(\hat{k}z)
    \sum\limits_{n=0}^{N/2} a_{2n}^c\cos(2nx)
    \ , \nonumber \\
    B &=& \varepsilon \cos(\hat{k}z)\sum\limits_{n=0}^{N/2} b_{2n+1}^c\cos((2n+1)x)\ .
    \label{16}
\end{eqnarray}
In this equation, $a^c_{2n}$ and $b^c_{2n+1}$ are the coefficients from Eq.\,(\ref{13}) for the unstable S1 mode. These coefficients were normalized so that the largest of them equals one in absolute value (the linear problem does not define the amplitude of the unstable disturbance). All computations were done with the small value of $\varepsilon = 0.01$ in Eq.\,(\ref{16}). Nonlinear properties of the instability therefore emerge after sufficient time $t \approx -\ln (\varepsilon )/\gamma$. The wave number  $\hat{k}$ in Eq.\,(\ref{16}) corresponds to the largest growth rate $\gamma$ in Fig.\,\ref{f2}.

The initial condition (\ref{16}) is periodic in either coordinate $x$ and $z$. The magnetic field remains periodic afterwards. The problem was therefore solved in a rectangle area of $-\pi \leq x \leq \pi$ and $-\pi/\hat{k} \leq z \leq \pi/\hat{k}$ with periodic boundary conditions.

As the instability is expected to enhance the field dissipation, it may be reasonable to follow the dynamics of the total magnetic energy,
\begin{equation}
    E = \frac{\hat{k}}{2\pi^2}\int\limits_{-\pi}^\pi
    \int\limits_{-\pi/\hat{k}}^{\pi/\hat{k}}
    \left(B^2 + \left(\frac{\partial A}{\partial x}\right)^2 +
    \left(\frac{\partial A}{\partial z}\right)^2\right)
    \mathrm{d}z\mathrm{d}x
    \label{17}
\end{equation}
(normalized to its initial value for the background field), and the power $W$ of the Ohmic dissipation,
\begin{eqnarray}
    W &=& \frac{\hat{k}}{2\pi^2}\int\limits_{-\pi}^\pi
    \int\limits_{-\pi/\hat{k}}^{\pi/\hat{k}}
    D(x,z) \mathrm{d}z\mathrm{d}x\ ,
    \nonumber \\
    D(x,z) &=& \left(\Delta A\right)^2 +
    \left(\frac{\partial B}{\partial x}\right)^2 +
    \left(\frac{\partial B}{\partial z}\right)^2
    \label{18}
\end{eqnarray}
(normalized similarly). In Eq.\,(\ref{18}), $D = j^2$ is the dissipation density whose spatial distribution can also be of some interest in relation with the possible formation of the current sheets.

\begin{figure}[!b]\centering
\includegraphics[width=\linewidth]{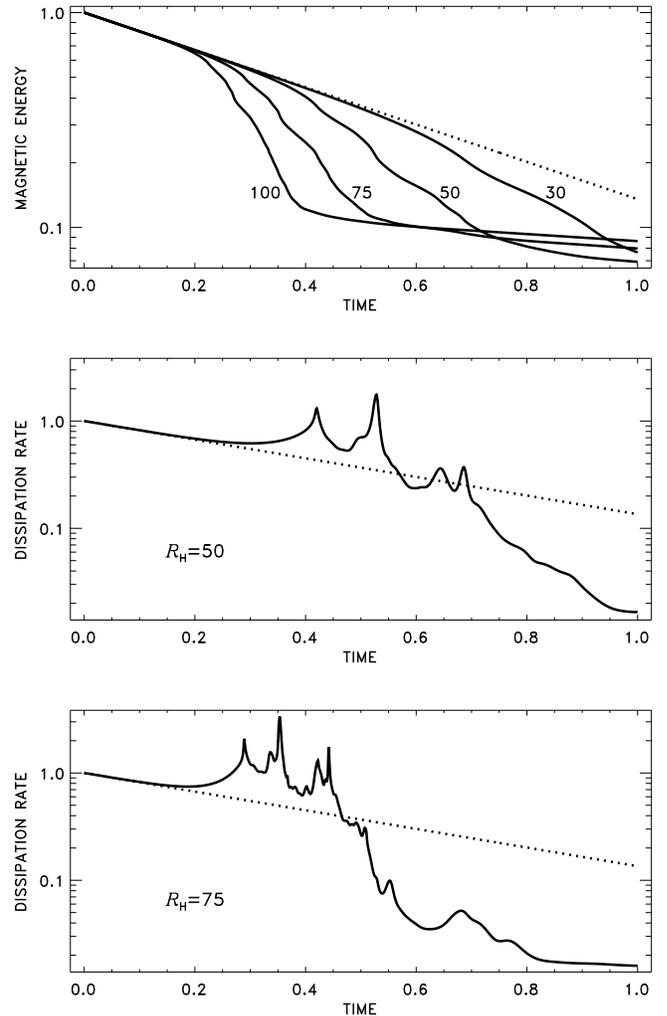}
 \caption{{\sl Top panel:} the dependencies of the magnetic energy $E$ (\ref{17})
    on time. Different lines are marked by the corresponding values of the Hall parameter $R_\mathrm{H}$. The dashed line shows the dependence  $\mathrm{exp}(-2t)$ for $R_\mathrm{H} = 0$. The {\sl middle and bottom panels} show the rate $W$ (\ref{18}) of the Ohmic dissipation for the Hall parameters $R_\mathrm{H} = 50$ and $R_\mathrm{H} = 75$, respectively.
    }
 \label{f5}
\end{figure}

\begin{figure*}[!t]\centering
\includegraphics[width=15.5truecm]{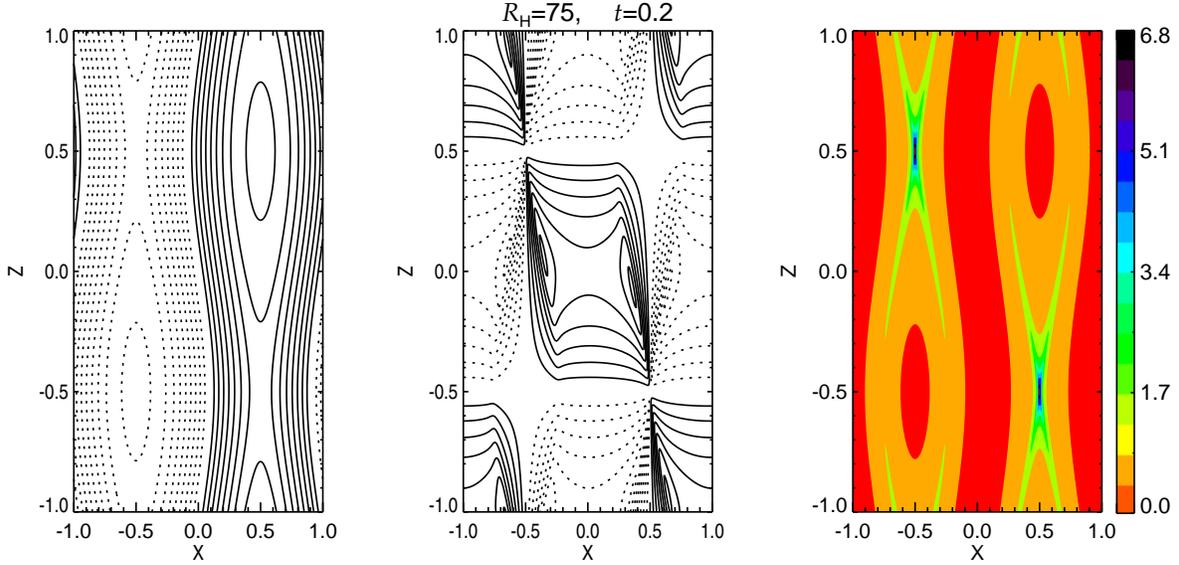}
 \caption{{\sl From left to right:} the poloidal field lines, contour lines
    of the toroidal field, and the pattern of the dissipation density $D$ (\ref{18}) for the instant $t = 0.2$ before the first spike of the energy release for $R_\mathrm{H} = 75$ in the Fig.\ref{f5}. Full and dotted lines show the positive and negative levels respectively of the potential $A$ (they coincide with the poloidal field lines) and the toroidal field $B$. The scale of the dissipation density $D$ is shown on the right. The same coordinates $X = \pi x$ and $Z = \pi\hat{k}z$ as in the Fig.\,\ref{f3} are used.
    }
 \label{f6}
\end{figure*}
\begin{figure*}[!t]\centering
\includegraphics[width=15.5truecm]{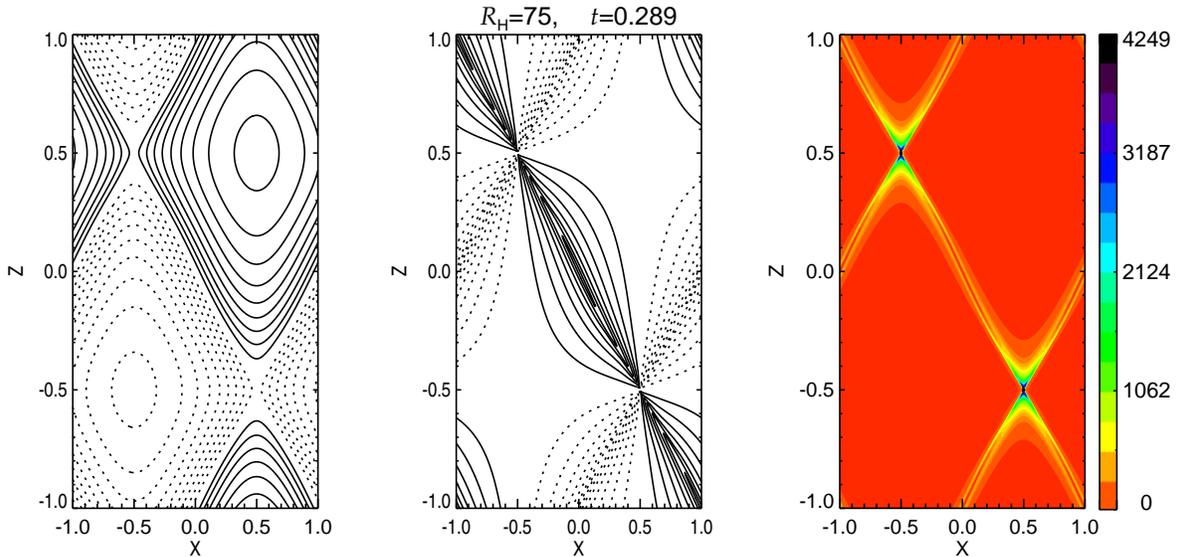}
 \caption{The same as in Fig.\,\ref{f6} but for the instant $t = 0.289$ of the first
    spike in the energy release.
    }
 \label{f7}
\end{figure*}
The problem was solved numerically using uniform finite-difference grids in $x$ and $z$. A numerical scheme with second-order accurate spatial derivatives and second-order Runge-Kutta time-stepping was applied.  The Hall term in Eq.\,(\ref{1}) as well as the diffusive term, includes spatial derivatives of the second order while the Hall parameter (\ref{12}) does not contain any spatial scale. Therefore, there is no spatial scale below which the field dynamics is dominated by diffusion. A decrease of scales due to the Hall effect is not balanced by diffusion and leads to rapid variations with time.  Numerical computations are very demanding to spatial and temporal resolution. The computations were initially performed with some number $N_x = N_z$ of the grid points. The computations were then repeated with about 1.5 times larger number of grid points. If the results were not distinguishable, the resolution was considered to be sufficient. Otherwise, the computations were repeated again with 1.5 times larger resolution. The resolution-inde\-pen\-de\-nt results were possible to obtain for moderate Hall parameters up to $R_\mathrm{H} = 75$. Some difference was noticeable between the results for $R_\mathrm{H} = 100$ obtained with 601 and 401 grid points in either dimension. Too slow computations for still higher resolution were not attempted.
%%%%%%%%%%%%%%%%%%%%%%%%%%%%%%%%%%%%%%%%%%%%%%%%%%%%%%%%%%%%%%%%%%%%%%%%%%%%%
\subsection{Results}
%%%%%%%%%%%%%%%%%%%%%%%%%%%%%%%%%%%%%%%%%%%%%%%%%%%%%%%%%%%%%%%%%%%%%%%%%%%%%
Figure \ref{f5} shows the time-dependencies of magnetic energy (\ref{17}) and dissipation rate $W$ (\ref{18}) for several values of the Hall parameter. The dashed lines in this figure show the decay law $\mathrm{exp}(-2t)$ without the Hall effect. Initially, all the computed trends follow these dashed lines closely. Considerable deviations develop, however, after time $t \approx -\ln(\varepsilon)/\gamma$ when the unstable disturbances amplitude becomes comparable with the background field. Such a nonlinear stage of  instability, obviously, onsets earlier for larger Hall parameters. The Hall effect does not change the magnetic energy by itself but the instability caused by this effect amplifies Ohmic dissipation.

Figure \ref{f5} shows that the increase in magnetic energy release is not steady but consists of a series of sharp increases or \lq spikes'. The spikes are related to the field dissipation in the current sheets. This can be seen by following the evolution of the magnetic field pattern.  Figure \ref{f6} shows the patterns of the field and the dissipation density $D$ (\ref{18}) for the instant $t = 0.2$ before the first spike, when the dissipation rate already starts increasing (all for $R_\mathrm{H} = 75$). The current sheets are clearly seen in this figure, especially in the dissipation density pattern of its right panel.

The magnetic energy release enhanced by the double Hall instability is highly inter\-mit\-tent not only in time but also in space. The current sheets formation took place in all nonlinear computations for not too small Hall parameters $R_\mathrm{H} \gsim 30$. The length of a current sheet decreases with time but the dissipation rate increases simultaneously. The rate attains its maximum value at the instant of the current sheet degeneration into an X-point (Fig.\,\ref{f7}). Note that the amplitude of the energy release density increases at this instant a thousand times (compared to its value for the background field), though the total dissipation power increases within one order of magnitude. The dissipation concentrates in a small vicinity of the X-points.

\begin{figure}\centering
\includegraphics[width=\linewidth]{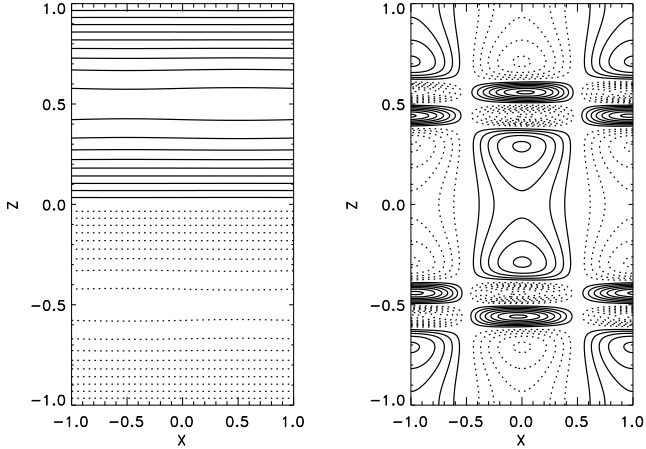}
 \caption{The stable field pattern for the final state $t = 1$ of the
    computation for $R_\mathrm{H} = 75$.
    }
 \label{f8}
\end{figure}

The X-point reconnection totally changes the field structure. Strictly speaking, the further field dynamics is not related to the original problem of field (\ref{9}) stability. However, the series of spikes with the energy release maxima at the instants of the current sheets degeneration into X-points continues afterwards. The series of reconnections terminates when the field distribution approaches a stable Hall equilibrium with a decre\-a\-s\-ed magnetic energy and a small dissipation rate. Figure \ref{f8} shows an example of such an equilibrium.
%%%%%%%%%%%%%%%%%%%%%%%%%%%%%%%%%%%%%%%%%%%%%%%%%%%%%%%%%%%%%%%%%%%%%%%%%%%%%
\section{Discussion}
%%%%%%%%%%%%%%%%%%%%%%%%%%%%%%%%%%%%%%%%%%%%%%%%%%%%%%%%%%%%%%%%%%%%%%%%%%%%%
The simplicity of our model and the restriction of the nonlinear computations by moderate Hall parameters allow only a qualitative comparison with the properties of pulsars. Goldreich \& Reisenegger (1992) estimated the Hall parameter for the neutron star crusts,
\begin{equation}
    R_\mathrm{H} \sim 400 \frac{B_{12}}{T_8^2}\left(\frac{\rho}{\rho_\mathrm{n}}\right)^2 ,
    \label{19}
\end{equation}
where $\rho_\mathrm{n} = 2.8\times10^{14}$ g/cm$^3$ is the \lq nuclear density' and the numerical subscripts mean, as usual, the order of magnitude: $B_{12}$ is the magnetic field in $10^{12}$\,G and $T_8$ is the temperature in $10^8$\,K. The parameter (\ref{19}) can be very large, especially for magnetars with immense magnetic fields $\sim 10^{15}$\,G. Double Hall instability can, therefore, take place if condition (\ref{8}) is satisfied. The $X$-ray luminosity of magnetars is in all probability supported by release of magnetic energy (Harding 2013). Radiation intensity is not uniform with time. It includes $\gamma$-ray bursts releasing enormous energy of $\gsim 10^{40}$\,erg in course of some tens of seconds. The bursts are qualitatively similar to the spikes of magnetic energy release of Fig.\,\ref{f5}.

Double Hall instability accelerates substantially the dissipation of magnetic energy. The field however relaxes to a stable state with low dissipation rate at the final stage of the nonlinear computations. This suggests a possible explanation for the fact that some of the rotation-powered pulsars are similar to the magnetars in the strength of their magnetic fields but do not show a comparably high magnetic activity (Harding 2013).

The similarity with shear-Hall instability suggests a pictorial explanation for the origin of double Hall instability (Fig.\,\ref{f1}) and allows estimation of its expected parameters. Linear stability analysis confirmed this estimations. The instability enhances substantially the resistive decay of the magnetic field at the nonlinear stage of its development.

Applications of the Hall effect are not restricted to pulsars. The effect is important for the stability of cool protostellar disks with a low degree of ionisation (R\"udiger \& Kitchatinov 2005) and for the solar corona (Stepanov et al. 2012). Reconnection in current sheets induced by the instability (Figs \ref{f6} and \ref{f7}) make the applications to the solar corona especially interesting. The Hall parameter for plasmas, $R_\mathrm{H} = \omega_\mathrm{e}/\nu_\mathrm{ei}$, is estimated by the ratio of the electron cyclotron frequency $\omega_\mathrm{e} = e B/(m_\mathrm{e} c)$ to the frequency $\nu_\mathrm{ei}$ of electron-ion collisions. An evaluation of the collision frequency for fully ionised Hydrogen (cf., e.g., Somov 2012) leads to the estimation
\begin{equation}
    R_\mathrm{H} \simeq 1.7\times 10^7\frac{B_2T_6^{3/2}}{n_9},
    \label{20}
\end{equation}
where $n_9$ is the electron density in $10^9$cm$^{-3}$. The large value of parameter (\ref{20}) indicates the significance of the Hall effect. Coronal magnetic fields are believed to be close to a force-free state, i.e., ${\vec\nabla}\times{\vec B} = \alpha_\mathrm{ff}{\vec B}$. The force-free fields, obviously, represent Hall equilibria. The equilibria can be unstable: condition (\ref{8}) for the double Hall instability is satisfied (it should be noted, however, that the condition was obtained for the simple case of a unidirectional field). An analysis of the instability for plasmas requires, nevertheless, separate consideration. Apart from the induction equation, such an  analysis should involve the motion equation and, probably, allow for anisotropy of transport coefficients. The conductivity of the neutron star crusts can be isotropic even for very strong magnetic fields (Urpin \& Shalybkov 1995), but for plasmas with large $R_\mathrm{H}$, the diffusion coefficients for the directions along and across the magnetic field can differ. A critical review of some experimental and theoretical results on reconnection in plasmas with the Hall effect is given by Somov (2013).
%%%%%%%%%%%%%%%%%%%%%%%%%%%%%%%%%%%%%%%%%%%%%%%%%%%%%%%%%%%%%%%%%%%%%%%%%%%%%
\phantomsection
\section*{Acknowledgments}
%\addcontentsline{toc}{section}{Acknowledgments} % Uncomment to add Acknowledgements to the table of contents
%%%%%%%%%%%%%%%%%%%%%%%%%%%%%%%%%%%%%%%%%%%%%%%%%%%%%%%%%%%%%%%%%%%%%%%%%%%%%
This work was supported by the Russian Foundation
for Basic Research (project 17--02--00016).
%%%%%%%%%%%%%%%%%%%%%%%%%%%%%%%%%%%%%%%%%%%%%%%%%%%%%%%%%%%%%%%%%%%%%%%%%%%%%
\phantomsection
\section*{References}
%\addcontentsline{toc}{section}{References} %Uncomment to add References to content
%%%%%%%%%%%%%%%%%%%%%%%%%%%%%%%%%%%%%%%%%%%%%%%%%%%%%%%%%%%%%%%%%%%%%%%%%%%%%
\begin{description}
%%%%%%%%%%%%%%%%%%%%%%%%%%%%%%%%%%%%%%%%%%%%%%%%%%%%%%%%%%%%%%%%%%%%%%%%%%%%%
\item{} Bisnovatyi-Kogan,~G.\,S., \& Komberg,~B.\,V. 1974, SvA {\bf 18}, 217
\item{} Bisnovatyi-Kogan,~G.\,S. 2016, arXiv:1601.04006
\newpage
\item{} Cumming,~A., Arras,~P., \& Zweibel,~E. 2004, \apj\ {\bf 609}, 999
\item{} Goldreich,~P., \& Reisenegger,~A. 1992, \apj\ {\bf 395}, 250
\item{} Gourgouliatos,~K.\,N., \& Cumming,~A. 2015, \mnras\ {\bf 446}, 1121
\item{} Gourgouliatos,~K.\,N., \& Hollerbach,~R. 2016, \mnras\ {\bf 463}, 3381
\item{} Harding,~A.\,K. 2013, Frontiers of Physics {\bf 8}, 679
\item{} Heyl,~J.\,S., \& Kulkarni,~S.\,R. 1998, \apj\ {\bf 506}, L61
\item{} Hollerbach,~R., \& R\"udiger,~G. 2002, \mnras\ {\bf 337}, 216
\item{} Jones,~P.\,B. 1988, \mnras\ {\bf 233}, 875
\item{} Kojima,~Y., \& Kisaka,~Y. 2012, \mnras\ {\bf 421}, 2722
\item{} Marchant,~P., Reisenegger,~A., Valdivia,~J., \& Hoyos,~J.\,H. 2014, \apj\ {\bf 796}, 94
\item{} Pons,~J.\,A., \& Geppert,~U. 2007, \aap\ {\bf 470}, 303
\item{} Pons,~J.\,A., \& Geppert,~U. 2010, \aap\ {\bf 513}, L12
\item{} Rheinhardt,~M., \& Geppert,~U. 2002, Phys. Rev. Lett. {\bf 88}, 101103
\item{} Reisenegger,~A., Benguria,~R., Prieto,~J.\,P., Araya,~P.\,A., \& Lai,~D. 2007, \aap\ {\bf 472}, 233
\item{} R\"udiger,~G., \& Hollerbach,~R. 2004, The magnetic universe (Weinheim: WILEY-VCH), p.296
\item{} R\"udiger,~G., \& Kitchatinov,~L.\,L. 2005, \aap\ {\bf 434}, 629
\item{} Shalybkov,~D.\,A., \& Urpin,~V.\,A. 1997, \aap\ {\bf 321}, 685
\item{} Somov,~B.\,V. 2012 Plasma Astrophysics, Part I (New York: Springer), \S\,8.1
\item{} Somov,~B.\,V. 2013 Plasma Astrophysics, Part II (New York: Springer) \S\,2.4.
\item{} Stepanov,~A.\,V., Zaitsev,~V.\,V., \& Nakariakov,~V.\,M. 2012, Physics Uspekhi {\bf 55}, A04
\item{} Urpin,~V., \& Geppert,~U. 1995, \mnras\ {\bf 275}, 1117
\item{} Urpin,~V.\,A., \& Shalybkov,~D.\,A. 1995, Astron. Rep. {\bf 39}, 332
\item{} Urpin,~V., \&  Shalybkov,~D. 1999, \mnras\ {\bf 304}, 451
\item{} Vainshtein,~S.\,I., Chitre,~S.\,M., \& Olinto,~A.\,V. 2000, Phys. Rev. E. {\bf 61}, 4422
\end{description}
\end{document}